\documentclass[12pt]{iopart}
\usepackage{epsfig}

\def\beq{\begin{equation}}
\def\eeq{\end{equation}}
\def\bea{\begin{eqnarray}}
\def\eea{\end{eqnarray}}

\def\ms {\overline{\rm MS}}

\def\d  {{\rm d}}
\def\gev{{\rm GeV}}

\def\lr {\left(   }
\def\rr {\right)  }
\def\le {\left[   }
\def\re {\right]  }

\begin{document}

\title[The pion structure function and jet production in $\gamma p\to nX$]{The
pion structure function and jet production in $\gamma p\to nX$}

\author{Michael Klasen
\footnote[3]{klasen@mail.desy.de}
}

\address{II.\ Institut f\"ur Theoretische Physik, Universit\"at Hamburg,
Luruper Chaussee 149, D-22761 Hamburg, Germany}

\begin{abstract}

Despite its theoretical and practical importance, the pion structure is still
badly constrained, particularly at low $x_{\pi}$ and in the sea-quark and gluon
sectors. Recently ZEUS have presented data on dijet photoproduction with
a leading neutron, which is dominated by slightly off-shell pion exchange
and can be used to constrain the pion densities down to $x_{\pi}\simeq 0.01$.
We compare a recent NLO calculation to the ZEUS data and find that the lower
gluon densities of SMRS seem to be preferred by the data. Theoretical
uncertainties, in particular from the pion flux, are discussed in some detail.

\end{abstract}



\maketitle

\section{Introduction}
\label{sec:1}

As one the simplest QCD bound states and as the Goldstone boson of chiral
symmetry breaking, the pion is a very intersting theoretical object: Its
structure carries important implications for the QCD confinement mechanism and
the realization of symmetries like isospin in nature. It is also of practical
importance for the hadronic input to the photon structure at low scales. The
latter is connected via Vector Meson Dominance to the $\rho$ meson structure,
which is poorly known and thus often replaced by the pion structure.

Unfortunately, determinations of the pion structure have made little progress
over the last decade. They are based on old Drell-Yan and prompt photon data
at fixed target energies and large values of the partonic momentum fraction
$x_{\pi}$. Many details are still based on pure theoretical assumptions.
In order to improve the situation, it has been proposed to measure the
(virtual) pion structure at low $x_{\pi}$ in deep inelastic scattering (DIS)
and photoproduction with leading neutrons at HERA \cite{Holtmann:1994rs}.
Since the pion is by far the lightest hadron, its exchange dominates the
$p\to n$ transition and it will almost be on its mass shell, particularly at
small values of the squared momentum transfer $t'$ between the proton and the
neutron.

Recently the H1 and ZEUS collaborations at HERA have installed forward proton
and neutron calorimeters and presented first results on the semi-inclusive DIS
structure functions $F_2^{\rm LP(3)}(\beta,x_p,Q^2)$ and $F_2^{\rm LN(3)}
(\beta,x_n,Q^2)$ \cite{Adloff:1999yg,zeusf2ln} and on dijet photoproduction
\cite{h1dijets,Breitweg:2000nk}. The latter has recently been compared to a
next-to-leading order (NLO) calculation \cite{Klasen:2001sg}.
In Sec.\ \ref{sec:2} of this paper, we define
the dijet photoproduction cross section, the kinematic variables, and the
experimental conditions of the scattering process. In Sec.\ \ref{sec:3} we
discuss the pion flux and in Sec.\ \ref{sec:4} the pion structure. The main
results are contained in Sec.\ \ref{sec:5}, where we compare our NLO dijet
calculation with the ZEUS data. A brief summary is given in Sec.\ \ref{sec:6}.

\section{Kinematics}
\label{sec:2}

\begin{figure}
 \begin{center}
  \epsfig{file=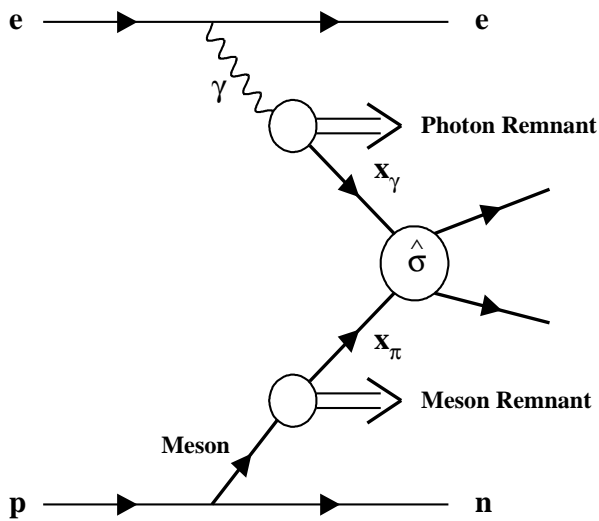,width=0.5\textwidth}
 \end{center}
 \caption{\label{fig:kinematics}
 Schematic diagram for dijet production with a leading neutron in
 photoproduction at HERA.}
\end{figure}

The production of two jets with a leading neutron at HERA is shown
schematically in Fig.\ \ref{fig:kinematics}.
The corresponding cross section
\beq
   \frac{\mbox{d}^3\sigma}{\mbox{d}E_T^2\mbox{d}\eta_1\mbox{d}\eta_2}
   = \sum_{a,b} x_a           f_{a/e}(x_a          ,M_\gamma^2) 
                  x_b           f_{b/p}(x_b          ,M_\pi^2   )
                  \frac{\mbox{d}\hat{\sigma}_{ab\to 12}}{\mbox{d}t}
\eeq
depends on the partonic cross section $\mbox{d}\hat{\sigma}_{ab\to 12}/\mbox{d}
t$ and on the parton densities in the electron and proton
\bea
  f_{a/e}(x_a,M_\gamma^2) &=& \int_{x_a}^1\frac{\d y}{y    }f_{a/\gamma}
   (x_\gamma,M_\gamma^2)f_{\gamma/e}(y,Q^2), \\
  f_{b/p}(x_b,M_\pi ^2  ) &=& \int_{x_b}^1\frac{\d (1-x_n)}{1-x_n}f_{b/\pi   }
   (x_\pi,M_\pi^2)f_{\pi/p}(1-x_n,t').
\eea
Both are convolutions of flux factors and parton densities. The photon flux
$f_{\gamma/e}(y,Q^2)$ is given by the usual Weizs\"acker-Williams
approximation \cite{vonWeizsacker:1934sx}.
The ranges of the photon momentum fraction $y=(P\cdot q)/(P\cdot k)
\in [0.2;0.8]$ and
virtuality $Q^2=-q^2<4\,\gev^2$ are dictated by the ZEUS experiment. The
momentum fraction of the leading neutron $x_n$ and the momentum transfer $t'$
in the pion flux factor $f_{\pi/p}(1-x_n,t')$ are restricted to $x_n=(P'\cdot
k)/(P\cdot k)> 400\,\gev\,/\,820\,\gev$ and $t'=f(p_T)=
-{p_T^2\over x_n}-{(1-x_n)(m_n^2-x_n
m_p^2)\over x_n}$ with $p_T<x_n\cdot 0.66\,\gev$. The hard jets are defined
with the $k_T$ cluster algorithm and a maximal distance
$R_{ij} = \sqrt{(\eta_i-\eta_j)^2+(\phi_i-\phi_j)^2} < 1$
\cite{Ellis:1993tq}. They are required
to have transverse energies above $E_T^{\rm jet} > 6\,{\rm GeV}$ and lie in the
central rapidity range $-2<\eta^{\rm jet}<2$.

\begin{figure}[b]
 \begin{center}
  \epsfig{file=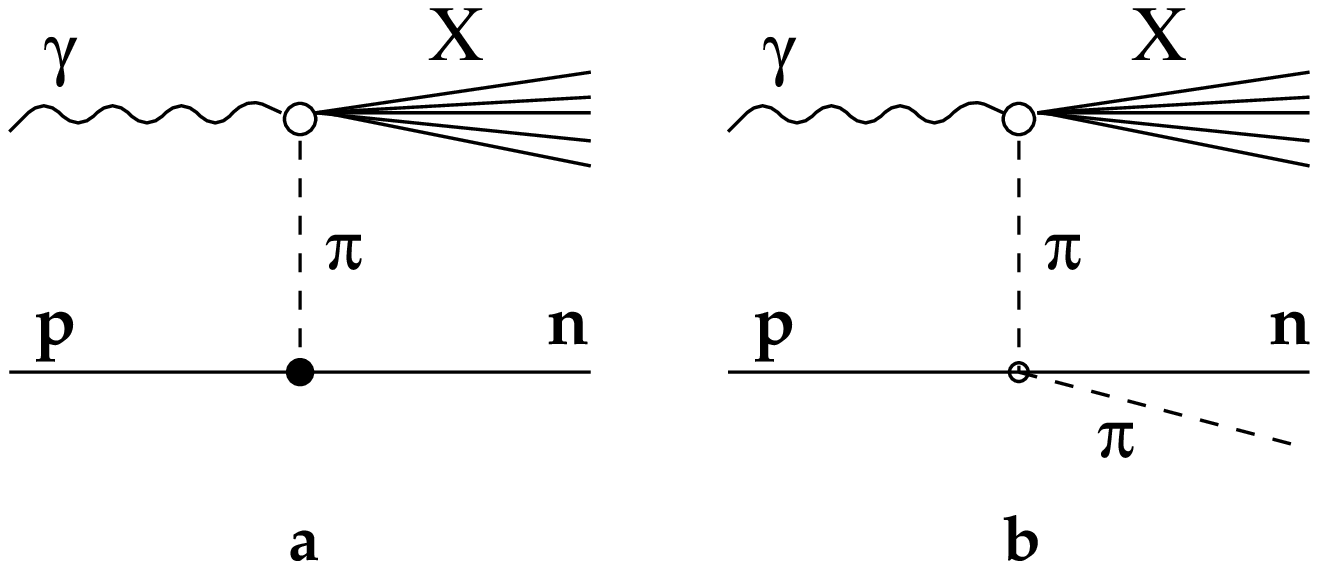,bbllx=113pt,bblly=312pt,bburx=496pt,bbury=482pt,%
          width=0.495\textwidth}
  \epsfig{file=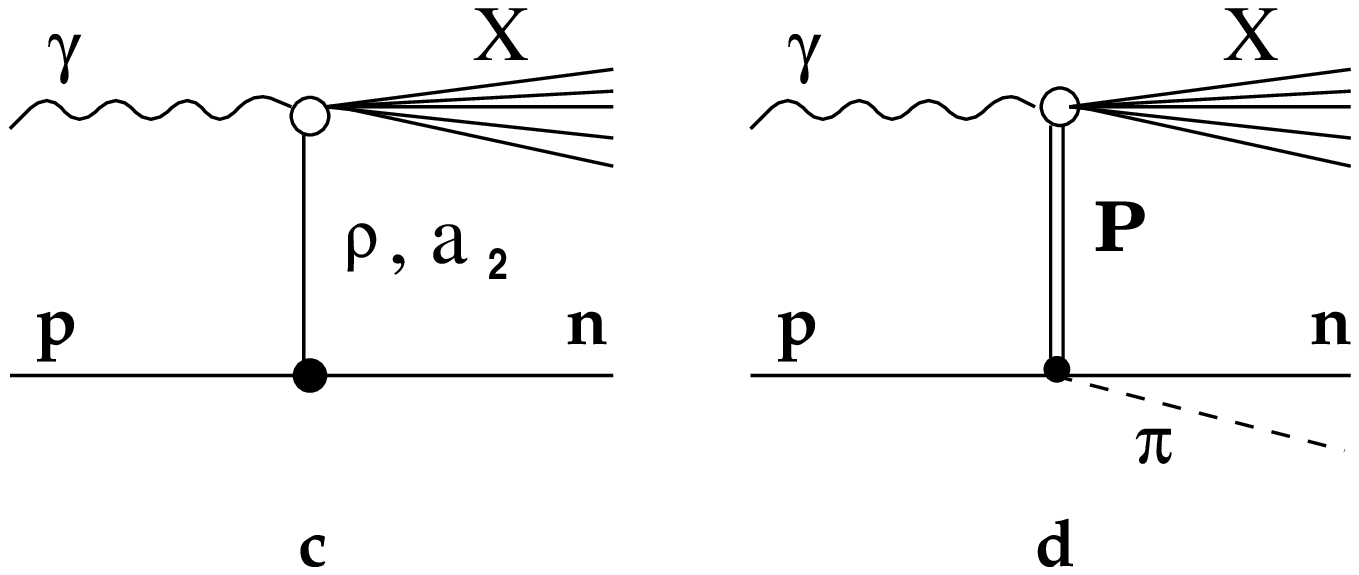,bbllx=113pt,bblly=312pt,bburx=496pt,bbury=482pt,%
          width=0.495\textwidth}
 \end{center}
 \caption{\label{fig:ope}
 Photoproduction with a leading neutron: a) Charged pion exchange, b)
 intermediate $\Delta$ production, c) Reggeon exchange, and d) Pomeron
 exchange.}
\end{figure}
\section{Pion Flux}
\label{sec:3}

In principle, a variety of soft reactions can lead to a leading neutron final
state at HERA. Four different possibilities are shown in Fig.\ \ref{fig:ope}:
a) charged pion exchange, b) intermediate $\Delta$ production, c) Reggeon
($\rho,\,a_2$) exchange, and d) Pomeron exchange.
Since the momentum transfer is restricted to low values $t'<0.54\,\gev^2$
in the ZEUS experiment, the pion as the lightest particle in the
$t'$-channel dominates.
Diagrams, where an additional pion is produced in the proton fragmentation
region (diagrams b) and d)), are experimentally excluded.
Reggeon exchange amounts to a small ($<10\,\%$) cross section increase.

\begin{figure}
 \begin{center}
  \epsfig{file=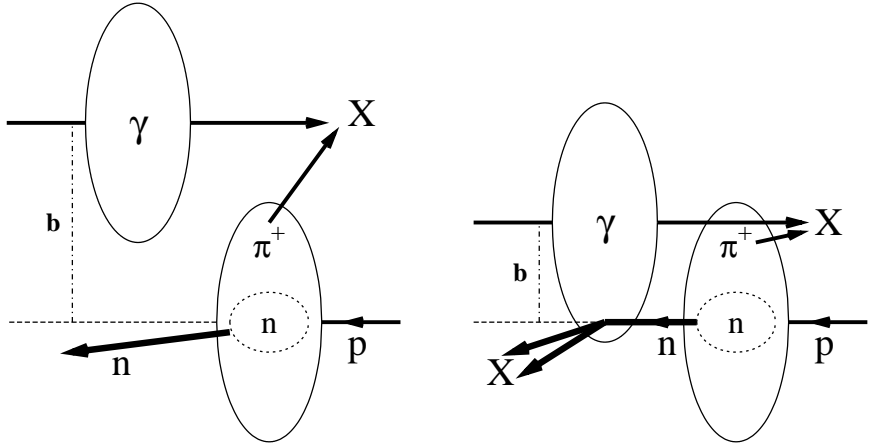,width=0.75\textwidth}
 \end{center}
 \caption{\label{fig:glaureac}
 Photoproduction with a leading neutron in the meson cloud model at large
 (left) and small (right) impact parameter $b$.}
\end{figure}

In the meson cloud model, the photon-proton collision can be viewed as a
process, where a pionic cloud is stripped off a neutron core (see Fig.\
\ref{fig:glaureac}).
As the impact parameter $b$ becomes small, additional effects like photon
absorption or neutron-rescattering may become important.
The semi-inclusive ZEUS data show that absorption effects decrease the cross
section by at most $10\,\%$ in the $x_n$ range considered, which cancels
approximately the increase of the cross section by Reggeon exchange.

The light cone representation of the proton wave function is given by
\beq\hspace*{-2.5cm}
 |p\rangle = \sqrt{Z}\le|p_0\rangle+\int\d x \d^2{\bf p_T}
 \lr \phi_{\pi/p}(x,{\bf p_T})\sqrt{1\over 3}|p,\pi^0\rangle
 +   \phi_{\pi/p}(x,{\bf p_T})\sqrt{2\over 3}|n,\pi^+\rangle
 + ...\rr\re,
\eeq
where $|p_0\rangle$ is the bare proton wave function, given by a three-quark
state, and $\phi_{M/p}(x,{\bf p_T})$ is the probability amplitude of finding a
baryon B (=p,n) with longitudinal momentum fraction $x$ and transverse
momentum ${\bf p_T}$ and a meson M (=$\pi^0,\pi^+$) in the proton. $Z$ is the
wave function renormalization constant and is fixed by the normalization
condition $\langle p|p\rangle=1$. The splitting function or flux factor takes
the general form
\bea
 f_{\pi/p}(1-x_n,t')&=& \int \d^2{\bf p_T}|\phi_{\pi/p}(x_n,{\bf p_T})|^2
 \delta(t'-f(p_T)) \\
 &=& \frac{3C_n}{4\pi}\frac{g_{n\pi p}^2}{4\pi}
  \frac{-t'}{(m_{\pi}^2-t')^2}(1-x_n)^{1-2\alpha_{\pi}'(t'-m_\pi^2)}
  [F(x_n,t')]^2.
\eea
It can be calculated in covariant or time-ordered perturbation theory.
$C_n$ is the squared Clebsch-Gordon isospin coefficient (2/3 for leading
neutrons, 1/3 for leading protons), $1/(m_{\pi}^2-t')^2$ is the squared pion
propagator, and $(1-x_n)^{1-2\alpha_{\pi}'(t'-m_\pi^2)}$ accounts for the
virtuality and possible reggeization of the pion. The interaction term in the
pion-nucleon Lagrangean leads to the numerator $-t'$, and off-mass shell
effects of the higher $|p,\pi^0\rangle$ and $|n,\pi^+\rangle$ Fock states
are modeled by a form factor
\beq
   F(x_n,t) = \left\{ \begin{array} {l} \exp [b(t-m_{\pi}^2)]  
  {\rm \hspace*{47mm} [Exponential]} \\ 
                                        \exp [R^2(t-m_{\pi}^2)/(1-x_n)]
  {\rm \hspace*{30mm} [Light Cone]}
                      \end{array} \right.
\eeq
The flux factor is constrained by a symmetry relation ($f_{\pi/p}(1-x_n,t')
=f_{n/p}(x_n,t')$) and charge and momentum sum rules.
Since the momentum transfer $t'$ is small, we neglect reggeization
($\alpha'=0$) and choose a light cone form factor with $R=0.5\,\gev^{-1}$
in good agreement with recent determinations
\cite{Holtmann:1996be,D'Alesio:2000bf}. The pion-nucleon coupling constant
$g^2/(4\pi)=14.17$ is taken from a recent extraction from the
Goldberger-Miyazawa-Oehme sum rule \cite{Ericson:2000md}.

\section{Pion Structure}
\label{sec:4}

\begin{figure}
 \begin{center}
  \epsfig{file=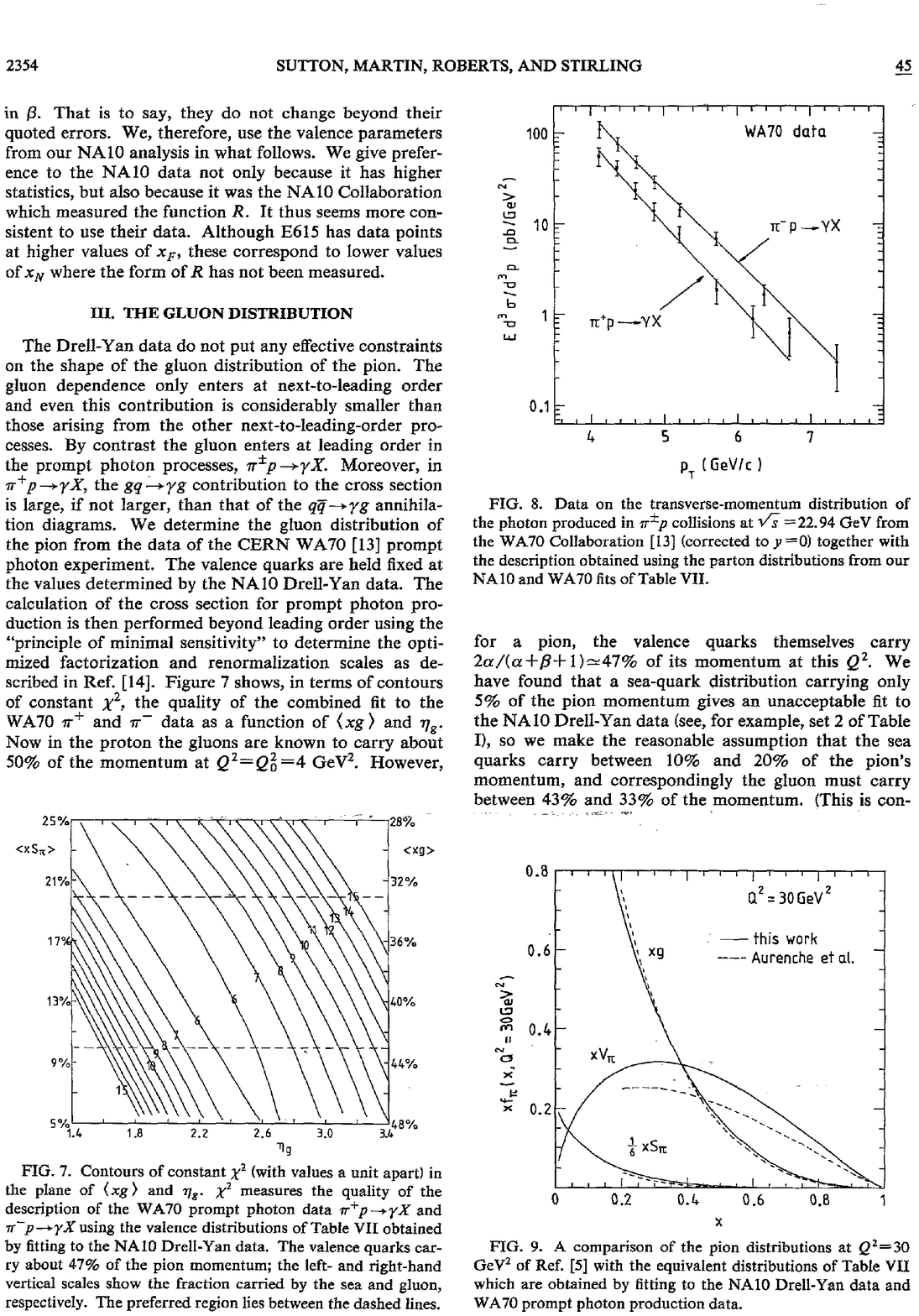,bbllx=73pt,bblly=129pt,bburx=311pt,bbury=320pt,%
          width=0.75\textwidth,clip=}
 \end{center}
 \caption{\label{fig:seagluon}
 Momentum fractions carried by sea-quarks and gluons in the pion as determined
 in the SMRS fit. The lowest $\chi^2$ values (6-15) are obtained with sea
 quark momentum fractions between 10 and 20 \%.~\cite{Sutton:1992ay}}
\end{figure}

\begin{figure}
 \begin{center}
  \epsfig{file=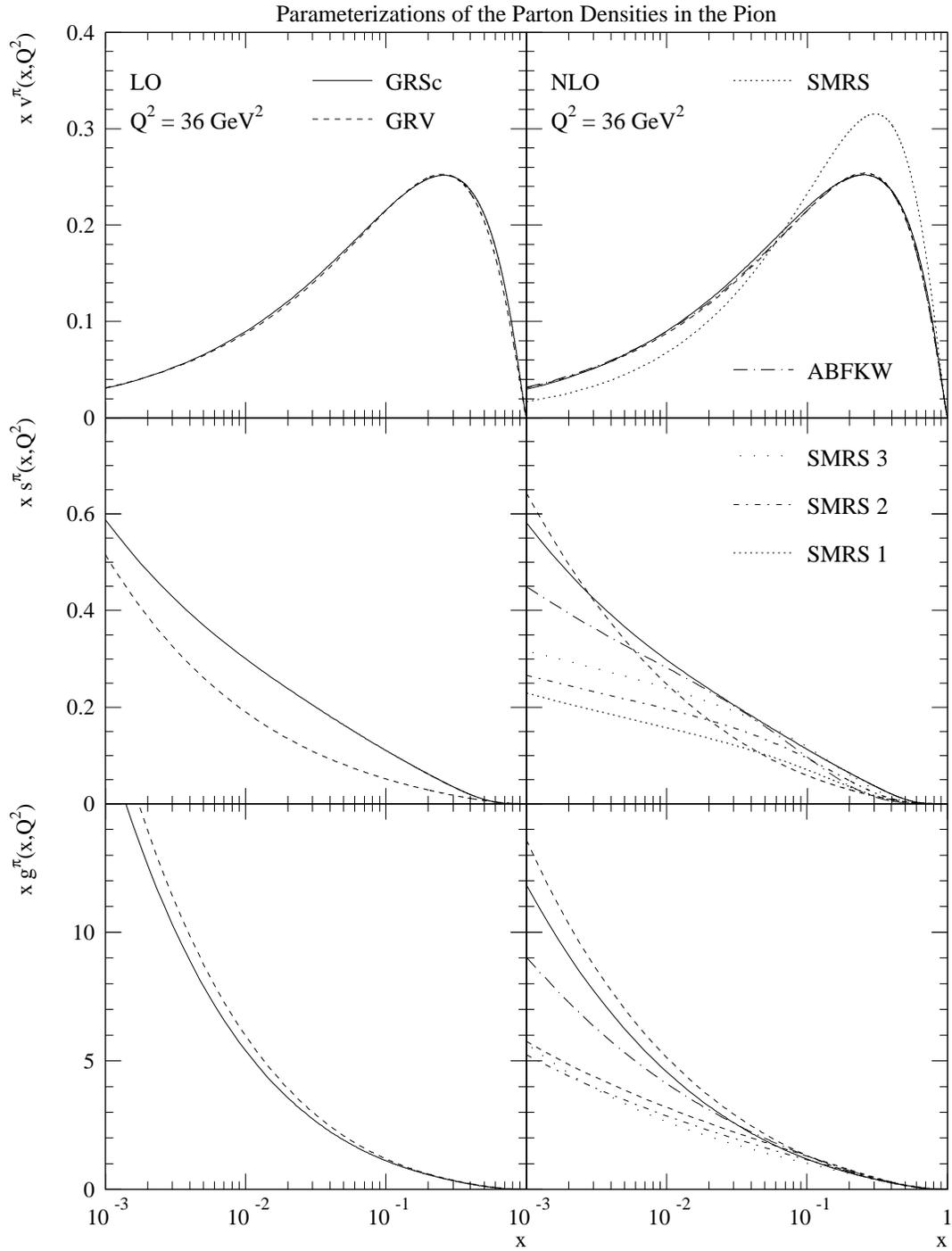,width=\textwidth}
 \end{center}
\caption{\label{fig:piqgy}
 Parameterizations of the valence-quark (top), sea-quark (center), and
 gluon (bottom) densities in the pion in LO (left) and NLO (right) at $Q^2=36$
 GeV$^2$.}
\end{figure}

Traditionally, the valence quark densities in the pion have been determined in
the Drell-Yan process, while the gluon density has been constrained in prompt
photon production (see Tab.\ 1).
\begin{table}
\begin{tabular}{|ccc|ccllcl|}
\hline
 Group & Year & Set & $Q_0^2$   & Fac.\ & Model          & Data & $N_f$ & $\Lambda_{\ms}^{N_f=4}$ \\
       &      &     & (GeV$^2$) & Sch.\   &                &      &       & (MeV)                   \\
\hline
\hline
 ABFKW & 1989 & NLO & 2.00      & $\ms$    & $v^\pi\,=$ $\gamma X$, DY     & WA70,NA24  & 4     & 229               \\
       &      &     &           &          & $s^\pi\,=$ DY                 & NA3        &       &                         \\
       &      &     &           &          & $g^\pi\,=$ $\gamma X$, MSR    & WA70,NA24  &       &                         \\
&&&&&&&&\\
  SMRS & 1992 & 10\%& 4.00      & $\ms$    & $v^\pi\,=$ DY                 & NA10, E615 & 4     & 190                     \\
       &      & 15\%&           &          & $s^\pi\,=$ DY                 & NA3        &       &                         \\
       &      & 20\%&           &          & $g^\pi\,=$ $\gamma X$, MSR    & WA70       &       &                         \\
&&&&&&&&\\
   GRV & 1992 & LO  & 0.25      & LO       & $v^\pi\,=$ ABFKW              & WA70,NA24  & 6     & 200                     \\
       &      & NLO & 0.30      & $\ms$    & $s^\pi\,=\,0$                 &            &       &                         \\
       &      &     &           &          & $g^\pi\,=$ MSR                &            &       &                         \\
&&&&&&&&\\
   GRSc& 1999 & LO  & 0.26      & LO       & $v^\pi\,=$ DY, MSR            & NA10, E615 & 3     & 204                     \\
       &      & NLO & 0.40      & $\ms$    & $s^\pi\,=\,(v^\pi/v^p)\, s^p$ & H1, ZEUS   &       & 299                     \\
       &      &     &           &          & $g^\pi\,=\,(v^\pi/v^p)\, g^p$ & H1, ZEUS   &       &                         \\
\hline
\end{tabular}
\caption{Parameterizations of the parton densities in the pion. They are
constrained by Drell-Yan (DY) and prompt photon production ($\gamma X$) data
and by the Momentum Sum Rule (MSR).}
\end{table}
In both cases the relevant data sets were taken in fixed target collisions in
the 1980s and are restricted to large $x_{\pi}\geq 0.2$. Furthermore, they
suffer from nuclear effects and fragmentation contributions, respectively,
and also from scale uncertainties.
The situation is worst for the sea quarks where theoretical assumptions vary
most widely. It is known that at $Q^2=4\,\gev^2$ valence quarks carry
about half of the pion's momentum. At the same scale gluons carry about half
of a proton's momentum. If pions and protons had a similar structure, as
assumed by GRSc \cite{Gluck:1999xe}, no room
would be left for the sea quarks. Indeed GRV assumed a zero-input density
for sea quarks, however at a very low scale \cite{Gluck:1992ey}. SMRS
determined that sea quarks should however carry between 10 and 20 \% of the
pion's momentum (see Fig.\ \ref{fig:seagluon}) \cite{Sutton:1992ay}.
In Fig.\ \ref{fig:piqgy} we compare the available parton densities in the
pion at a scale $Q^2=E_{T,\min}^2=36\,\gev^2$ relevant for dijet
photoproduction on a slightly virtual pion. As one can see, the SMRS sea
quarks and gluons turn out lower than the very similar ABFKW, GRV, and GRSc
distributions \cite{Aurenche:1989sx}.

\section{Dijet Production}
\label{sec:5}

The ZEUS collaboration have compared their dijet results with a leading
neutron with LO Monte Carlo predictions and found reasonable agreement.
It is, however, well-known that pure LO results are strongly scale dependent,
that they can be modified by potentially large NLO corrections, and that they
do not allow for the implementation of a jet algorithm.
For inclusive dijet photoproduction, four different NLO calculations have been
performed using different phase space slicing methods and the subtraction
method. They all agree well within numerical errors \cite{Harris:1998ss}.
We have implemented in our calculation \cite{Klasen:1997it} two additional
integrations, {\it i.e.} over $x_n$
and $t'$, and the pion flux factor. The proton parton densities have been
replaced with those in the pion. 

\begin{figure}
 \begin{center}
  \epsfig{file=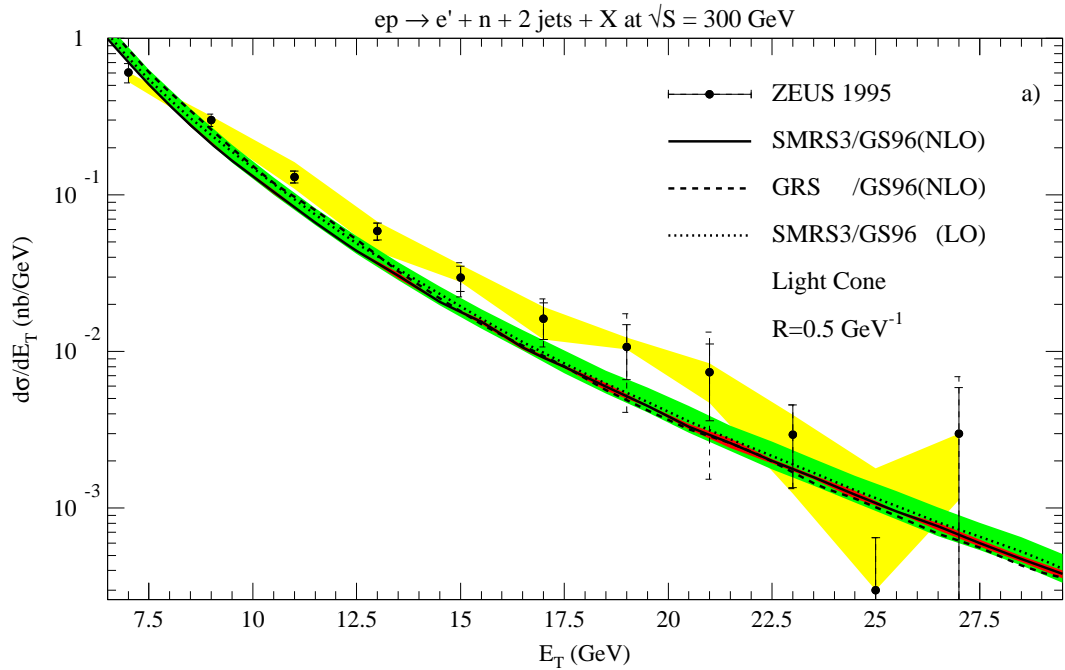,width=\textwidth}
 \end{center}
 \caption{\label{fig:1a}
 Dependence of the dijet photoproduction cross section with a leading neutron
 on the transverse energy of one of the two jets. Two NLO predictions with
 different pion structure functions and one LO prediction are shown together
 with experimental data from ZEUS with statistical (inner) and systematic
 (outer) experimental error bars. In
 addition, error bands from the theoretical scale uncertainty (medium: LO,
 dark: NLO) and the experimental energy scale uncertainty (light) are shown.
 The NLO error band coincides with the thickness of the full line.}
\end{figure}

Selecting the two jets with highest $E_T$ and integrating over the $\eta\in[-2;
2]$, we compare in Fig.\ \ref{fig:1a} the LO and NLO $E_T$ distribution with
GS96 photon densities \cite{Gordon:1997pm} and two different pion densities to
the ZEUS data. Almost identical results are obtained with GRV photon
densities.~\cite{Gluck:1992jc}
The $K$-factor turns out to be one, but the scale dependence is reduced
considerably from LO to NLO. The normalization is sensitive to the chosen
pion flux factor: If one includes the Regge trajectory ($\alpha'=1\,\gev^{-2}$)
and omits the exponential, the cross section is reduced by $15\%$. The same
reduction is obtained with a light cone form factor and $R=0.6\,\gev^{-1}$.
Resolved photons and gluons in the pion are important at low $E_T<9\,\gev$ and 
$E_T<22\,\gev$, respectively.
\begin{figure}
 \begin{center}
  \epsfig{file=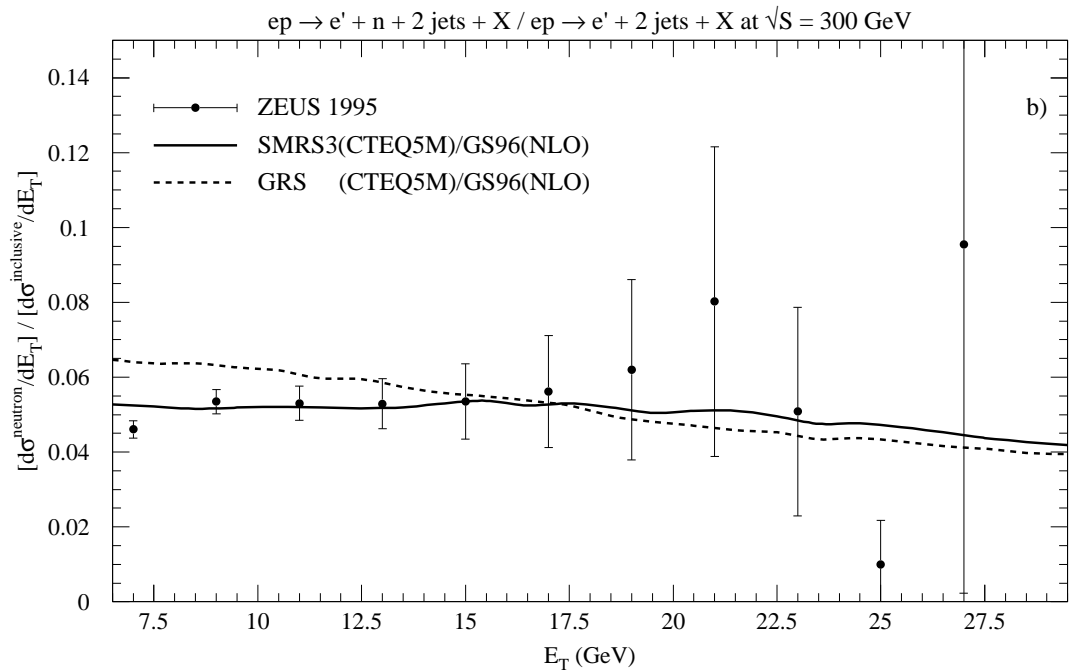,width=\textwidth}
 \end{center}
 \caption{\label{fig:1b}
 Ratio of the leading neutron over the inclusive
 cross section versus the transverse energy. NLO calculations with two
 different pion structure functions are shown together with the experimental
 ZEUS data. Only statistical error bars are given.}
\end{figure}
In the ratio of leading neutron and inclusive dijet cross sections in Fig.\
\ref{fig:1b}, many uncertainties from the scale choice, photon densities, and
hadronic energy scale cancel. The $E_T$ shape is then sensitive to the pion
densities and shows a preference for SMRS3 over GRSc.

\begin{figure}
 \begin{center}
  \epsfig{file=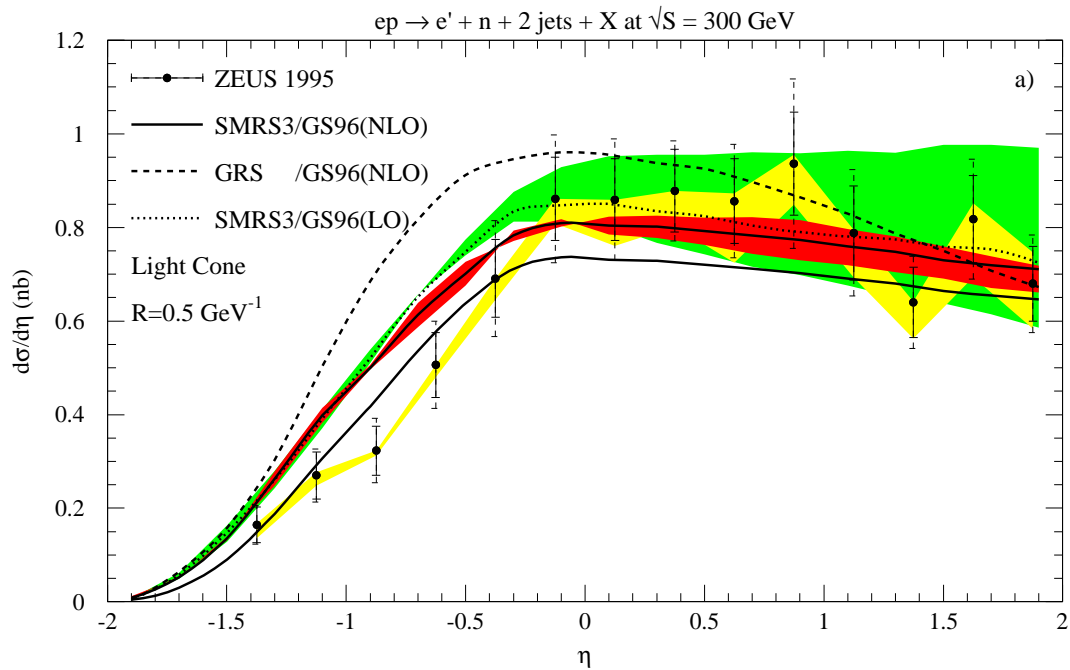,width=\textwidth}
 \end{center}
 \caption{\label{fig:2a}
 Dependence of the dijet photoproduction cross section with a leading neutron
 on the rapidity of one of the two jets. Details are as in Fig.\ \ref{fig:1a}.
 The lower full curve demonstrates the influence of
 hadronization corrections.}
\end{figure}

\begin{figure}
 \begin{center}
  \epsfig{file=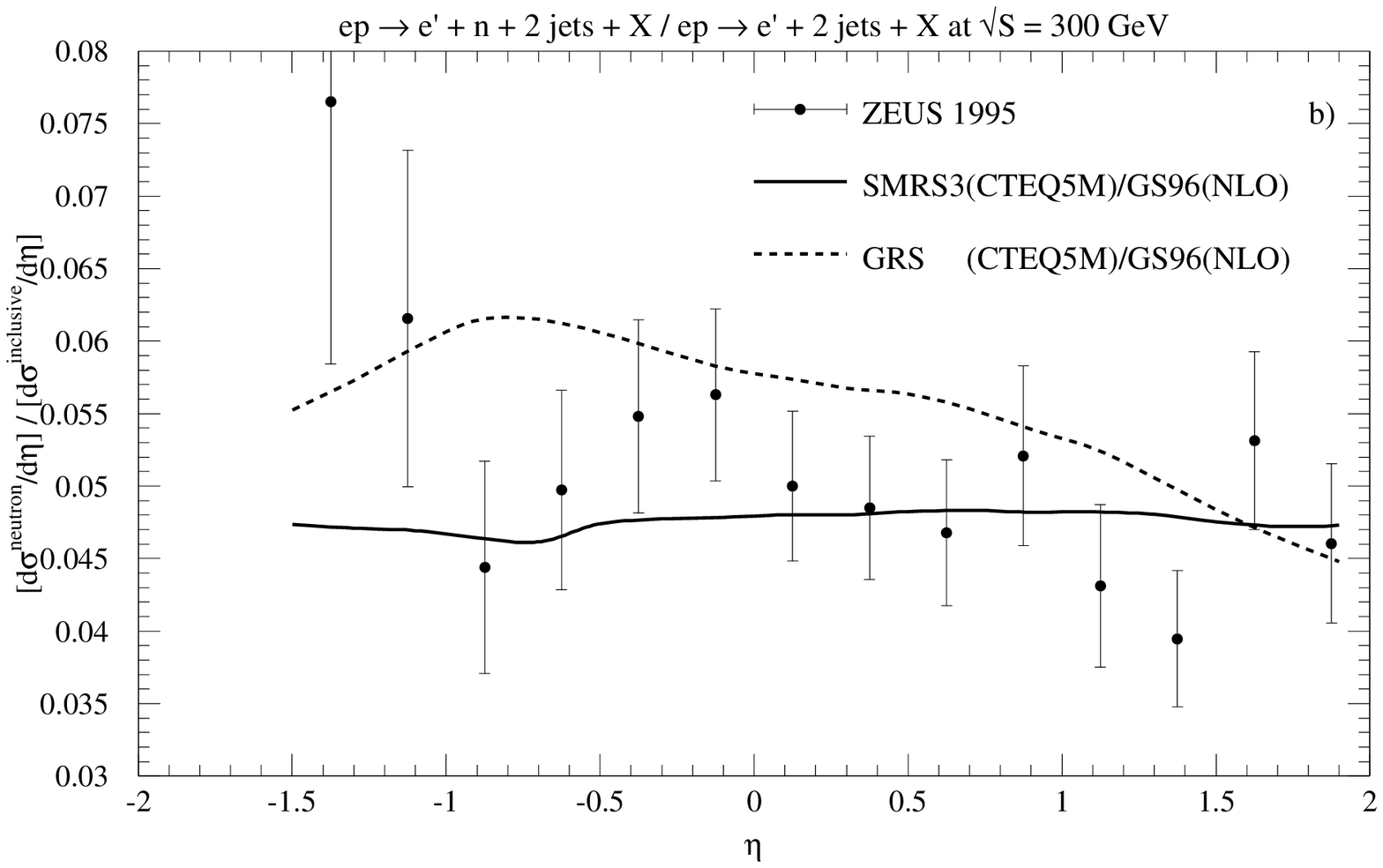,width=\textwidth}
 \end{center}
 \caption{\label{fig:2b}
 Ratio of the
 leading neutron over the inclusive cross section versus the rapidity. Details
 are as in Fig.\ \ref{fig:1b}.}
\end{figure}

If the leading neutron cross section is integrated over the two highest jet
$E_T>6\,\gev$, but not over $\eta$, one obtains the rapidity distribution 
shown in Fig.\ \ref{fig:2a}. Direct photons and quarks in the photon are
important at $\eta<0$. In this region of jet photoproduction, the low quarks
in the NLO GS96 parameterization usually give the best fit, despite the fact
that they do not reproduce the photon structure function.~\cite{Vogt:1997jt}
Indeed GS96 also gives the best fit for dijets with a leading neutron.
At $\eta<0$ hadronization corrections become also important and can amount to a
factor of two.
The data for the ratio of leading neutron over inclusive rapidity
distributions in Fig.\ \ref{fig:2b} have still rather large statistical 
error bars and do not distinguish between different pion parameterizations.

\begin{figure}
 \begin{center}
  \epsfig{file=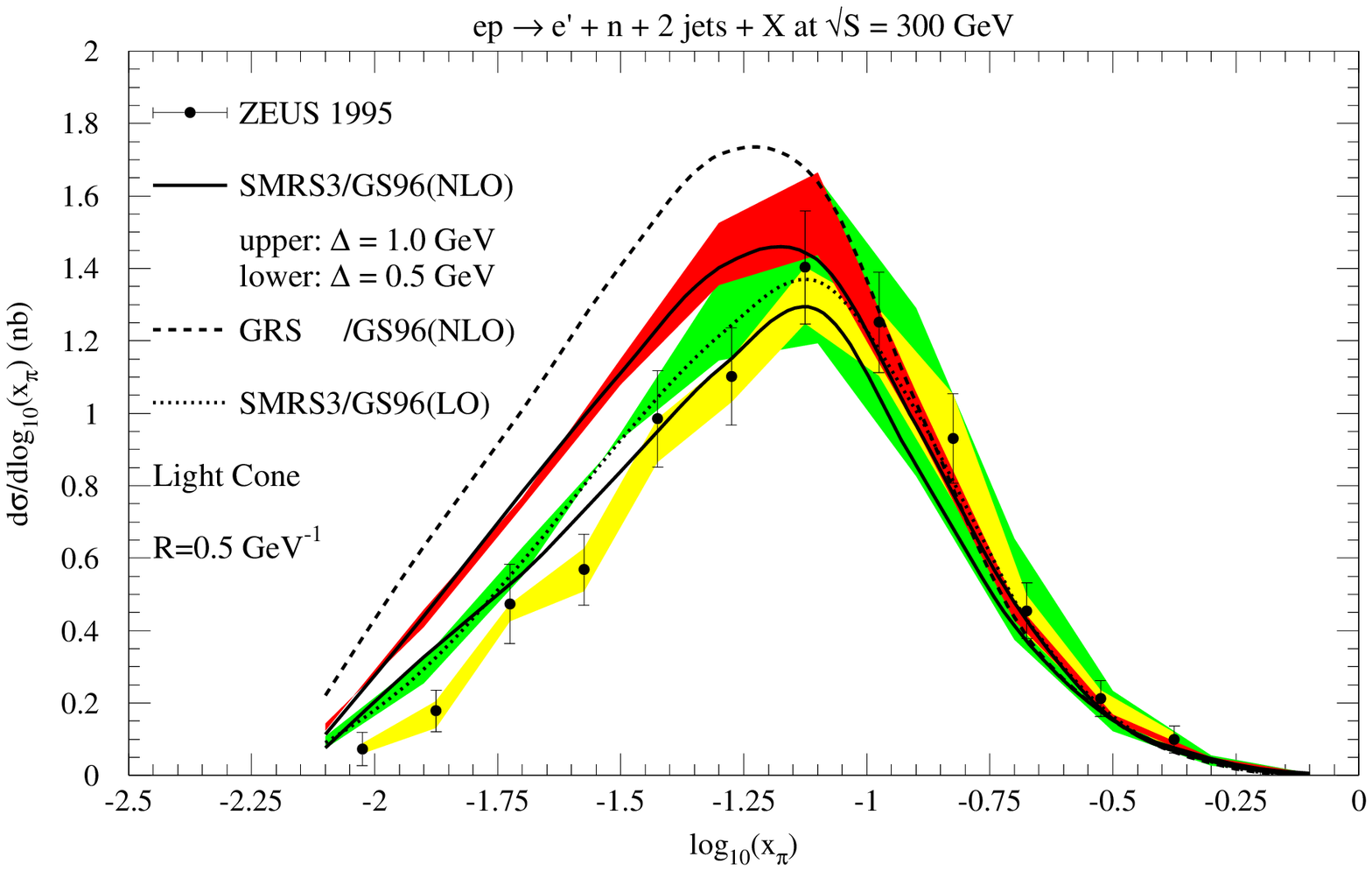,width=\textwidth}
 \end{center}
 \caption{\label{fig:3}
 Dependence of the dijet photoproduction cross section with a leading neutron
 on the logarithm of the observed momentum fraction of the partons in the pion
 $\log_{10}(x_\pi)$. Details are as in Figs.\ \ref{fig:1a} and \ref{fig:2a}.
 In addition, we show
 the effect of lowering the transverse energy difference $\Delta$ from 1 GeV
 to 0.5 GeV.}
\end{figure}

The distribution in the observed momentum fraction of the partons in the
pion
\beq
 x_{\pi}^{\rm OBS} = \frac{E_{T_1}e^{ \eta_1}+E_{T_2}e^{ \eta_2}}
 {2 (1-x_n)E_p}
\eeq
is shown in Fig.\ \ref{fig:3}. It suffers from the fact that ZEUS have
applied equal $E_T$ cuts on both hard jets, which is known to be infrared
sensitive \cite{Klasen:1996xe}. Therefore the cut on the second jet in the
NLO calculation has to be relaxed to $E_{T,2}>5$ or 5.5 GeV.
The best fit is again obtained with GS96 photon densities and SMRS3 pion
densities, which have the lowest quark and gluon distributions, respectively.
It should be kept in mind that the quark densities in the photon are correlated
with the gluon density in the pion.
In the $x_{\pi}$-distribution, hadronization corrections may play an important
role at small $x_{\pi}$, which is related to negative rapidities.

\section{Summary}
\label{sec:6}

Despite their theoretical and practical importance, the parton densities in the
pion are still poorly constrained, particularly below $x_{\pi}<0.2$.
An intriguing possibility for improvement consists in dijet photoproduction
with a leading neutron at HERA, which is dominated by slightly off-shell
pion exchange. Here the pion can be probed down to $x_{\pi}\simeq 0.01$.
However, theoretical uncertainties from Reggeon exchange and photon absorption,
modeled by a pion flux factor, have to be well under control.

Data on dijet photoproduction with a leading neutron have been obtained by
the ZEUS collaboration. Unfortunately, ZEUS have applied theoretically
unsafe $E_T$ cuts. Furthermore, the data suffer from large hadronization
corrections in the backward region. On the other hand, the hard scattering
cross section has been calculated in NLO and is theoretically stable.
A comparison of the NLO calculation with data indicates that lower gluon
densities in the pion, as predicted by the SMRS3 parameterization, are
preferred over steeper gluons as predicted by ABFKW, GRV, or GRSc.

\section*{Acknowledgments}

I wish to thank the organizers of the Ringberg Workshop for the kind
invitation, A.~Vogt for useful discussions, and G.~Kramer for his
collaboration. Financial support by the
Deutsche Forschungsgemeinschaft through Grant No.\ KL~1266/1-1, by the
Bundesministerium f\"ur Bildung und Forschung through Grant No.\ 05~HT9GUA~3,
and by the European Commission through the Research Training Network
{\it Quantum Chromodynamics and the Deep Structure of Elementary Particles}
under Contract No.\ ERBFMRX-CT98-0194 is gratefully acknowledged.

\end{document}